\documentclass[twocolumn,prl]{revtex4}

\usepackage{graphicx}
\usepackage{dcolumn}
\usepackage{bm}
\usepackage{textcomp}

\begin{document}

\title{Strain phase diagram and domain orientation in SrTiO$_3$ thin films}

\author{Feizhou He}
\author{B. O. Wells}%
\affiliation{Department of Physics, University of Connecticut,
Storrs, Connecticut 06269, USA}%

\author{S. M. Shapiro}
\affiliation{Department of Physics, Brookhaven National
Laboratory, Upton, New York 11973, USA}%

\date{\today}

\begin{abstract}
SrTiO$_3$ thin films were used as a model system to study the
effects of strain and epitaxial constraint on structural phase
transitions of perovskite films. The basic phenomena revealed will
apply to a variety of important structural transitions including
the ferroelectric transition. Highly strained SrTiO$_3$ films were
grown on different substrates, providing both compressive and
tensile strain. The measured strain-temperature phase diagram is
qualitatively consistent with theory, however the increase in the
phase transition temperature is much larger than predicted. Due to
the epitaxial strain and substrate clamping, the SrTiO$_3$ lattice
is tetragonal at all temperatures. The phase transitions involve
only changes in internal symmetry. The low temperature phase under
tensile strain has a unique structure with orthorhombic $Cmcm$
space group but a tetragonal lattice, an interesting consequence
of epitaxial constraint.
\end{abstract}

\maketitle

An important area of research in recent years has been to
understand exactly why the properties of epitaxial films differ
from related bulk materials, and to learn how to use these
differences to engineer desirable properties. Effects due to
strain in films are often thought of as analogous to those of high
pressure experiments. The main differences are: (1) The stress in
films is typically biaxial rather than hydrostatic or uniaxial.
(2) Films are subject to the additional constraint from the
substrate. (3) Films can have much larger strains than is usually
achievable in pressure cells. (4) Films are easier to work with
for many experiments and applications. Ferroelectricity in
strained perovskite films has been a particularly noteworthy topic
of study. For example, recent experiments have shown
strain-induced ferroelectricity in SrTiO$_3$ (STO) films, and huge
changes in the ferroelectric transition temperature $T_C$ in both
SrTiO$_3$ and BaTiO$_3$ (BTO) films under strain
\cite{Haeni04,Choi04}. Other notable results include large changes
in the metal-insulator transition temperature of RNiO$_3$ films
under strain \cite{Tiwari02} and reports of increasing the
transition temperature of La$_{2-x}$Sr$_{x}$CuO$_4$ under
compressive strain \cite{Sato97,Locquet98,Si99,Bozovic02}.

In this letter, we present a systematic investigation of the
structural phase transition in epitaxial STO films with varying
degrees of both compressive and tensile strain. This transition is
not ferroelectric, though it has been described by the same theory
as used for the ferroelectric transition in STO \cite{Pertsev00}
and referenced in the works above. We construct a
strain-temperature phase diagram for this transition with several
data points over a wide range of strain. The observed enhancement
of the structural phase transition temperature $T_s$, is much
greater than predicted by theory, though in many aspects the
theoretical predictions are qualitatively correct. We also note
that in some cases the film structures have unique symmetry not
possible in a free bulk system.

We chose the anti-ferrodistortive structural phase transition in
STO as a model system for phase transitions in epitaxial films.
The principal advantage of this system is that both the primary
and secondary order parameters are directly accessible through
x-ray diffraction measurements \cite{He04}, as explained below.
This is in contrast to the ferroelectric phase transition in
materials such as BTO where the primary order parameter is the
polarization, which is best detected by electrical measurements.
However, these phase transitions are similar in many ways: the
variation of lattice constant is a secondary order parameter for
both, changes in critical temperature with strain vary in a
similar manner, and similar domain orientations are present in the
low symmetry phases.

Bulk STO crystal is cubic at room temperature, with space group
$Pm\bar{3}m$. Below 105 K it becomes tetragonal with $I4/mcm$
symmetry. This phase transition involves the rotation of TiO$_6$
octahedra, and the rotation angle has been identified as the order
parameter for this phase transition \cite{Muller68}. In terms of
the pseudo-cubic unit cell, the tetragonal phase has additional
superlattice peaks at half integer index positions. The
intensities of the superlattice peaks are proportional to the
square of the order parameter and can be used to track the phase
transition. A secondary order parameter is the tetragonality,
which in the bulk is coupled to the octahedra rotation.

For a second-order structural phase transition, usually the volume
of the unit cell changes smoothly through the transition
temperature, but lattice parameter versus temperature curves have
a sudden change in slope. This is the case for bulk STO
\cite{Lyt64,Ale69}. For epitaxial films the in-plane lattice
parameters are subject to lateral constraint from the substrate
and therefore do not have the freedom to change as in bulk. In
previous studies we have discussed the effect of strain and
substrate clamping on several film systems, noting that this
substrate clamping effect is often described in too simplistic a
manner \cite{He04,He03}.

The STO films studied were grown on four kinds of substrates:
SrLaAlO$_4$ (SLAO), LaAlO$_3$ (LAO),
(LaAlO$_3$)$_{0.3}$(Sr$_2$LaTaO$_6$)$_{0.7}$ (LSAT) and KTaO$_3$
(KTO). All substrates are $\langle$0 0 1$\rangle$ oriented single
crystals. The mismatch between STO and these substrates ranges
from -3.82\% for SLAO to +2.15\% for KTO. Pulsed laser deposition
(PLD) was used to grow the films. The energy density of the KrF
excimer laser was about 1J/cm$^2$. The films were grown at
720\textcelsius\ in 100 mTorr O$_2$, with thickness ranging from 7
nm to 90 nm. X-ray diffraction showed excellent epitaxy with
average mosaics around 0.1 degrees and no detectable misoriented
regions.

X-ray diffraction measurements were carried out at beamline X22A
at the National Synchrotron Light Source, Brookhaven National
Laboratory. X22A has a bent Si(1 1 1) monochromator, giving a
small beam spot and fixed incident photon energy of 10 keV. The
angular resolution with a graphite (0 0 2) analyzer was less than
0.006\textdegree\ FWHM for an (0 0 2) peak, as measured from the
LAO substrate. Below room temperature the sample was cooled in a
closed cycle refrigerator with a temperature control of
$\pm$0.5\textdegree.

\begin{table}
\caption{\label{tab:lat-par}Lattice parameters and in-plane
strains of STO films on different substrates. The in-plane lattice
parameters of substrates are: SLAO -- 0.3756 nm, LAO -- 0.3790 nm,
LSAT -- 0.3868 nm, and KTO -- 0.3989 nm. For bulk STO $a=0.3905$
nm.}
 \begin{ruledtabular}
 \begin{tabular}{cccccc}
 STO& &\multicolumn{3}{c}{STO Lattice Par. (nm)}&In-plane\\
 Thickness&Substrate&$a_1$&$a_2$&$c$&Strain\footnotemark[1]\\
 \hline
 7 nm&SLAO&\multicolumn{2}{c}{0.3887}&0.3921&-0.46\%\\
 35 nm&SLAO&\multicolumn{2}{c}{0.3886}&0.3934&-0.49\%\\
 10 nm&LAO&\multicolumn{2}{c}{0.3884}&0.3922&-0.54\%\\
 90 nm&LAO&\multicolumn{2}{c}{0.3888}&0.3935&-0.44\%\\
 7 nm&LSAT&\multicolumn{2}{c}{0.3864}&0.3924&-1.04\%\\
 35 nm&LSAT&\multicolumn{2}{c}{0.3871}&0.3951&-0.86\%\\
 \hline
 7 nm&KTO&0.3984&0.3878\footnotemark[2]&0.3984&+2.03\%\\
 20 nm&KTO&0.3971&0.3879\footnotemark[2]&0.3973&+1.72\%\\
\end{tabular}
\end{ruledtabular}
 \footnotetext[1]{Positive -- tensile strain, negative -- compressive strain.}
 \footnotetext[2]{$a_2$ axis is out-of-plane for
films under tensile strains.}
\end{table}

Table~I shows the lattice parameters of STO films on various
substrates, by measuring several Bragg peaks at room temperature.
The random measurement errors in lattice parameters are less than
0.00005~nm, though systematic errors may be larger. The in-plane
biaxial strain is defined as $\epsilon=(a_{||}-a_0)/a_0$, where
$a_0$ is the lattice parameter of bulk STO (0.3905 nm), and
$a_{||}$ is the average of the lattice parameters along two
in-plane axes in the strained films. On KTO substrate, we obtained
very large tensile strain in 7 nm STO films. The in-plane lattice
parameters are almost equal to KTO value, resulting in an in-plane
strain of +2.03\%. The out-of-plane lattice parameters shrinks to
0.3878 nm. The 20 nm sample shows partial relaxation, with
in-plane strain of +1.7\%.

The films on LSAT substrates show largest compressive strain. Due
to the small mismatch between STO and LSAT, even the 35 nm film is
pseudomorphic with LSAT. Unfortunately, LSAT exhibits
face-centered-cubic type ordering \cite{Li03}, which results in
additional peaks at half integer positions. This makes following
the STO superlattice peaks impossible. Because of the large
mismatch, even very thin films on SLAO or LAO substrates are
already partially relaxed. The residual compressive strain within
the STO layer is around -0.5\% for both substrates. The phase
transitions in 10 nm and 90 nm STO / LAO samples were
investigated.

The tetragonality, $(c/a)-1$, of the STO unit cell does not change
dramatically through the whole temperature range. This is
consistent with our previous observations \cite{He03,He04}. The
tetragonality in all our films is considerably larger than in
bulk.

\begin{figure}
\includegraphics[scale=1]{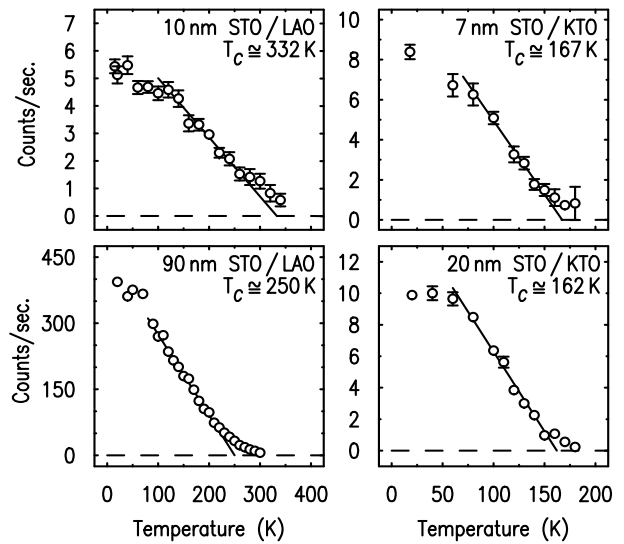}
\caption{\label{fig:sto-int}Temperature dependence of STO
superlattice peak intensities, showing $T_s$s for different
samples. For LAO (KTO) substrate, the strain in STO is compressive
(tensile).}
\end{figure}

Fig.~\ref{fig:sto-int} shows the temperature dependence of one of
the half-integer superlattice peaks associated with the low
temperature phase of STO. Results from several different films are
shown. The STO films also have a phase transition with similar
internal symmetry change as seen in the bulk. The transition
temperature is enhanced under conditions of both tensile and
compressive strain. A much larger enhancement is present for
compressive strain.

\begin{figure}
\includegraphics[scale=1]{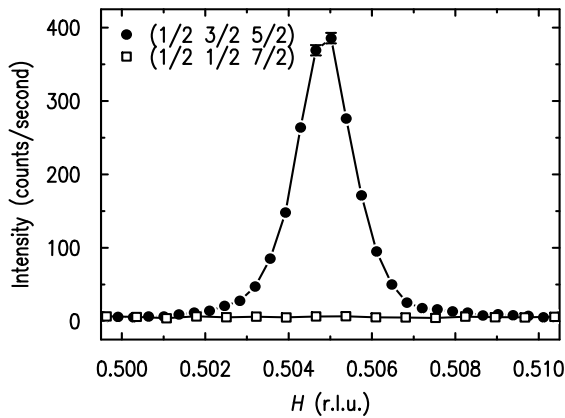}
\caption{\label{fig:sto-cc}The presence of (1/2 3/2 5/2) peak but
not (1/2 1/2 7/2) peak at 20 K in 90 nm STO / LAO sample proves
that the unique axis $c$ of STO film is out-of-plane.}
\end{figure}

The tetragonal bulk phase of STO (space group $I4/mcm$) is
characterized by the rotation of the TiO$_6$ octahedra around the
$c$-axis. The appropriate selection rules denote some peaks such
as (1/2 1/2 7/2) as forbidden, while (1/2 7/2 1/2) and (7/2 1/2
1/2) peaks are permitted. We can use the selection rules to
determine the domain orientation in the films. However, in the
following we continue to follow the standard practice of denoting
the direction normal to the surface as $\langle$0 0 1$\rangle$. We
refer to the unique tetragonal axis as merely the axis of
rotation. Fig.~\ref{fig:sto-cc} shows the relevant peaks for one
of the STO films on LAO, the standard case for compressive strain.
The (1/2 1/2 7/2) peak is missing. This indicates that the axis of
rotation coincides with the normal to the plane. This is not
surprising since in bulk STO, the axis of rotation is longer than
the others. And in the case of compressive strain, epitaxial
strain forces the out-of-plane axis to be longer than the in-plane
axes. Further, this means that the films are single domain and
have space group $I4/mcm$ as in the bulk. The high temperature
phase does differ from the bulk in that it is tetragonal rather
than cubic, thus having space group $P4/mmm$. The phase transition
in this case is from a high temperature, high symmetry tetragonal
phase ($P4/mmm$) to a low temperature, low symmetry tetragonal
phase ($I4/mcm$).

The films under tensile strain have a somewhat more interesting
morphology. All peaks of the type (1/2 1/2 7/2), (1/2 7/2 1/2),
and (7/2 1/2 1/2) are present. This indicates at least two domain
orientations. Due to the tensile strain, and the fact that the
rotation axis is the longer in the bulk, the most likely
orientations are with the rotation axis along each of the primary
in-plane directions. Thus there is a 90\textdegree\ twin
structure. As far as we can measure, the two in-plane lattice
constants remain identical regardless of the orientation of the
rotation axis. Thus we have a very unique morphology for the low
temperature phase. The most likely space group is orthorhombic
$Cmcm$. This space group is a subgroup of $I4/mcm$, but lacks the
4-fold axis. It also permits unequal dimensions along all three
directions. The lattice itself appears to be tetragonal in that $c
\approx a_1 \neq a_2$. This situation is fundamentally allowed
since the space group refers only to symmetry operations and not
the shape of the unit cell itself. The existence of a phase with
an orthorhombic space group but a tetragonal lattice is an
interesting consequence of strain and substrate clamping. We do
not believe that such a situation is possible in a free crystal.
In this case the phase transition is from the same high
temperature, high symmetry tetragonal phase ($P4/mmm$) to a low
temperature, orthorhombic phase ($Cmcm$).

The orientation of the unique axis $c$ below $T_s$ is quite
similar to the ferroelectric polarization in the recently reported
electric measurements on similar films \cite{Haeni04}. Similar
symmetry considerations will apply. It is also possible that in
STO the two phase transitions are intimately coupled, thus it will
be interesting to investigate the relationship between the two.

\begin{figure}
\includegraphics[scale=1]{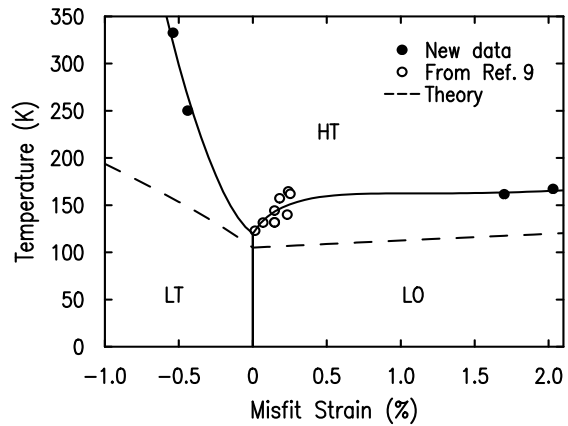}
\caption{\label{fig:phase-diag}Structural phase diagram of
strained STO. The theoretical prediction from
Ref.~\onlinecite{Pertsev00} is shown as dashed lines for
comparison. Solid lines are from a fitting to experimental data
\cite{He-up}.}
\end{figure}

Our data allows us to construct the strain phase diagram for STO
thin films. The new $T_s$ points reported in this letter extend
the range of strain considerably in both compressive and tensile
directions, as shown in fig.~\ref{fig:phase-diag}, where high
symmetry tetragonal, low symmetry tetragonal and orthorhombic
phases are labelled HT, LT and LO respectively. For tensile
strain, the experimental data show rapid increase of $T_s$ over
the small strain regime. Then $T_s$ stabilizes at about
160$\sim$170 K for larger strain up to 2\%. For compressive
strain, the $T_s$ increase rapidly even for relatively small
strain. A 0.5\% compressive strain results in a $T_s$ over 200 K
higher than bulk value, reaching room temperature region.

A theoretical investigation of this phase transition in epitaxial
films has been performed by Pertsev et al. \cite{Pertsev00}. Our
data corresponds to the structural part of their phase diagram,
designated by Pertsev as HT/SO and HT/ST transition, while
Ref.~\onlinecite{Haeni04} reflects the ferroelectric part. Our
result is qualitatively consistent with the theory in many
aspects. The domain structures we measured for both compressive
and tensile strain were predicted correctly. In addition, the
general trend in the variation of the transition temperature with
strain occurs much as predicted. The $T_s$ rises for both strains,
though compressive strain causes a much larger increase than does
tensile strain. The major difference between the data and theory
is that the magnitude of the rise in $T_s$ is much larger than
predicted. In addition, the shape of the actual $T_s$ versus
strain curves deviates from the nearly-linear prediction.

We have analyzed this result in terms of the theory of Pertsev et
al., where $T_s$ is calculated from the ralationship $\beta^*_i=0$
using equation (3) in Ref.~\onlinecite{Pertsev00}. Trying to fit
the data to this form but allowing the materials parameters to
float would result in unreasonable values of the elastic
constants. It is possible that a more complicated coupling term
must be included in a proper theory. However, we note that the
amount of strain in our films is much larger than is achieved in
measurements of bulk elastic constants. For such large strains,
the elastic response may no longer be linear. We would expect that
compression would become more difficult at large strains but
tension easier, due to the fundamental nature of atomic
interactions. It appears that such variation of the elastic
constants would allow for an expression similar to Pertsev's to
match our data \cite{He-up}, though a proper theory needs to be
constructed.

In conclusion, highly strained SrTiO$_3$ films were obtained on
different substrates. Under compressive strain, the phase
transitions in STO films are from high symmetry tetragonal to low
symmetry tetragonal, while under tensile strain, the transitions
are from high symmetry tetragonal to orthorhombic. The effects of
strain and substrate clamping induces structures in epitaxial
films that are not possible for bulk materials. The structural
phase transition temperature $T_s$ is enhanced by both compressive
strain and tensile strain. Many aspects of the strain-temperature
phase diagram are well described by current theory though
improvements are necessary to describe the magnitude of the
increase in transition temperature.

We acknowledge S. P. Alpay and B. Misirlioglu for helpful
discussion. This material is based upon work supported by the
National Science Foundation under Grant No. DMR-0239667 (BW, FH).
BW thanks the Cottrell Scholar Program of the research corporation
for partial support of this work. Work at Brookhaven is supported
by Division of Material Sciences, U.S. Department of Energy under
contract DE-AC02-98CH10886.


\end{document}